\documentstyle[twocolumn,prb,epsf,aps]{revtex}

\begin{document}
\draft
\title{An effective Hamiltonian for an extended Kondo lattice model and a possible
origin of charge ordering in half-doped manganites}
\author{Shun-Qing Shen and Z. D. Wang}
\address{Department of Physics, The University of Hong Kong, Pokfulam Road, Hong Kong}

\date{January 6, 1999}

\twocolumn[\hsize\textwidth\columnwidth\hsize\csname@twocolumnfalse\endcsname
\maketitle
\begin{abstract}
An effective Hamiltonian is derived in the case of the strong Hund coupling
and on-site Coulomb interaction by means of a projective perturbation
approach. A physical mechanism for charge ordering in half-doped manganites
(R$_{0.5}$X$_{0.5}$MnO$_3$) is proposed. The virtual process of electron
hopping results in antiferromagnetic superexchange and a repulsive
interaction, which may drive electrons to form a Wigner lattice. The phase
diagram of the ground state of the model is presented at half doping. In the
case of formation of Wigner lattice, we prove that spins of electrons are
aligned ferromagnetically as well as that the localized spin background is
antiferromagnetic. The influence of the on-site Coulomb interaction is also
discussed.
\end{abstract}


]


\section{Introduction}

The family of doped manganites, R$_{1-x}$X$_{x}$MnO$_{3}$ (where R = La, Pr,
Nd; X=Sr, Ca, Ba, Pb), has renewed both experimental and theoretical
interests due to the colossal magnetoresistance and its potential
technological application to magnetic storage devices. Apart from their
unusual magnetic transport properties, experimental observations of a series
of charge, magnetic and orbital ordering states in a wide range of dopant
also stimulate extensive theoretical curiosities. Early theoretical studies
of manganites concentrated their effort on the existence of metallic
ferromagnetism. From the so-called ``Double Exchange'' (DE) model, \cite
{Zener51} in which the mobility of itinerant electrons forces the localized
spins to align ferromagnetically, one can understand qualitatively the
relation of transport and magnetism. However the rich experimental phase
diagrams are far beyond the DE model. For example, according to the DE
model, itinerant electrons have the lowest kinetic energy in a tight binding
model, and should be driven to form a more stable ferromagnetic phase when
the system is half doped, {\it i.e.}, x=0.5. On the contrary, it is
insulating rather than metallic ferromagnetic at a low temperature as
expected theoretically. Furthermore, a charge ordered state was observed,
which is characterized by an alternating Mn$^{3+}$ and Mn$^{4+}$ ions
arrangement in the real space. \cite{Jirak85} Usually when the repulsive
interaction between charge carriers dominates over the kinetic energy the
charge carriers are driven to form a Wigner lattice. It has been shown
experimentally that the charge ordering is sensitive to an applied magnetic
field at low temperatures: resistance of a sample may decrease in several
order of magnitude and the charge ordering disappears at a low temperature, 
\cite{Tomioka95} which implies that the repulsive interaction should have a
close relation to the spin background. Although there have been extensive
theoretical efforts on anomalous magnetic properties, \cite{Millis95} a
comprehensive understanding on the physical origin of ordered states and
their relations to the transport properties are still awaited.

To explore electronic origin of these phenomena, we try to establish a more
unified picture to understand the physics starting from an electronic model,
which has been used to investigate the magnetic properties of the system
extensively. We derive an effective Hamiltonian in the case of the strong
on-site Coulomb interaction and Hund coupling by means of a projective
perturbation approach. It is found that the virtual process of electron
hopping produces an antiferromagnetic superexchange coupling between
localized spins and a repulsive interaction between itinerant electrons. The
antiferromagnetic correlation will enhance the repulsive interaction and
suppress the mobility of electrons. In the half-doped case, i.e., $x=0.5 $,
relatively strong repulsion will drive electrons to form a Wigner lattice.
In the case of the Wigner lattice, we prove that the electrons are fully
saturated while the localized spins form an antiferromagnetic background.
Strictly speaking, the ground state possesses both anti- and ferromagnetic,
i.e., ferrimagnetic long-range orders.

\section{Effective Hamiltonian}

The electronic model for doped manganites studied in this paper is defined
as \cite{Inoue95} 
\begin{eqnarray}
H &=&-t\sum_{\langle ij\rangle ,\sigma }c_{i,\sigma }^{\dagger }c_{j,\sigma
}+U\sum_{i}n_{i,\uparrow }n_{i,\downarrow }  \nonumber \\
&-&J_{H}\sum_{i}{\bf S}_{i}\cdot {\bf S}_{ic}+J_{AF}\sum_{\langle ij\rangle }%
{\bf S}_{i}\cdot {\bf S}_{j}.  \label{ele-ham}
\end{eqnarray}
where $c_{i,\sigma }^{\dagger }$ and $c_{i,\sigma }$ are the creation and
annihilation operators for $e_{g}$ electron at site $i$ with spin ${\sigma }$
$(=\uparrow ,\downarrow )$, respectively. $\langle ij\rangle $ runs over all
nearest neighbor pairs of lattice sites. ${\bf S}_{ic}=\sum_{\sigma ,\sigma
^{\prime }}{\bf \sigma }_{\sigma \sigma ^{\prime }}c_{i,\sigma }^{\dagger
}c_{i,\sigma ^{\prime }}/2$ and ${\bf \sigma }$ are the Pauli matrices. $%
{\bf S}_{i}$ is the spin operator of three $t_{2g}$ electrons with the
maximal value $3/2$. $J_{H}>0$ is the Hund coupling between the $e_{g}$ and $%
t_{2g}$ electrons. The antiferromagnetic coupling originates from the
virtual process of superexchange of $t_{2g}$ electrons. In reality, the $%
e_{g}$ orbital is doubly degenerated. For the sake of simplicity, we only
consider one orbital per site, which amounts to assuming a static
Jahn-Teller distortion and strong on-site interactions (relative to kinetic
energy).

\begin{figure}[tbp]
\epsfysize=6cm
\epsfxsize=8.5cm
\epsfbox{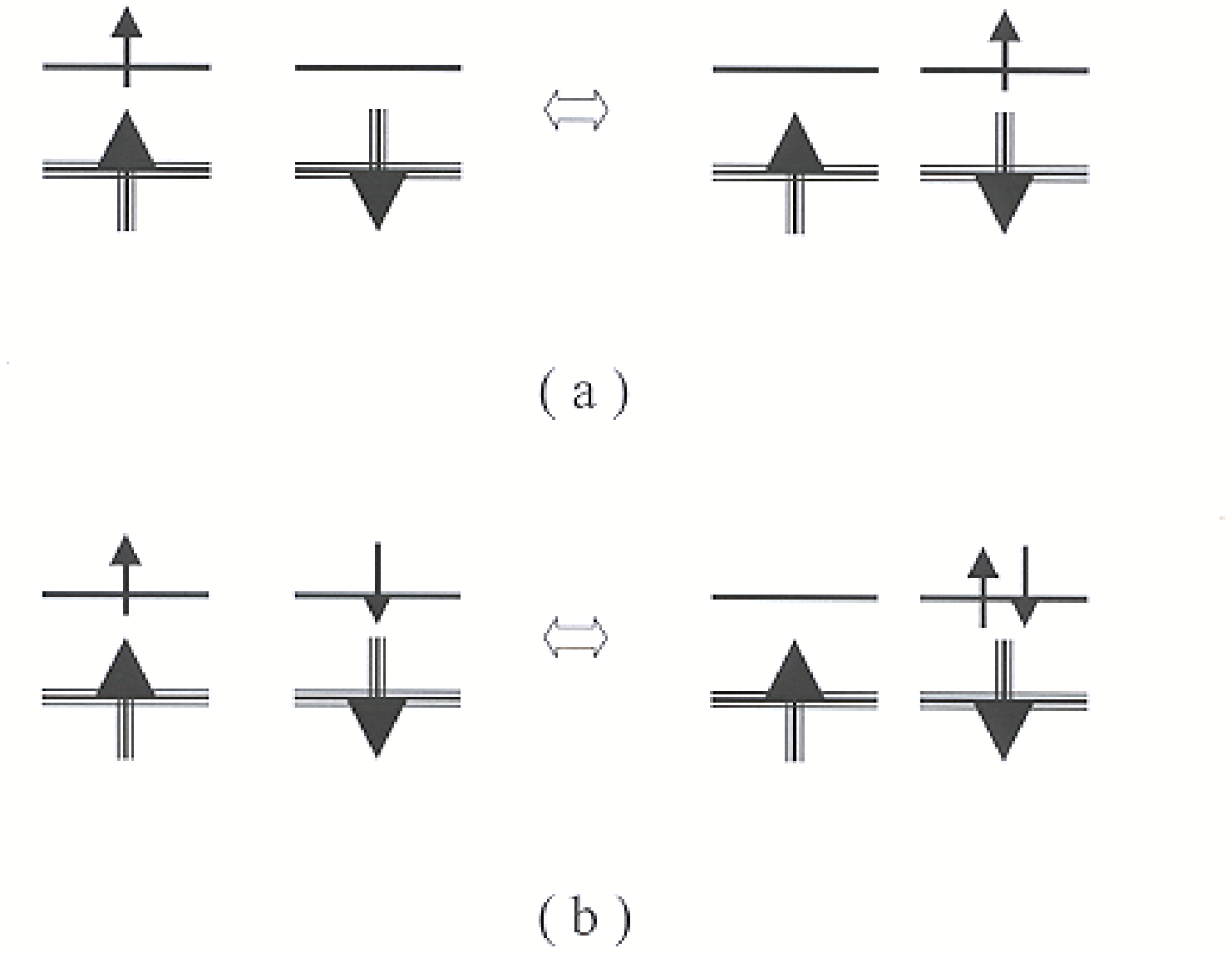}
\caption{ Tow virtual processes of electron hopping in the restricted
Hilbert space which favors antiferromagnetic correlation between neighboring
sites. On the left side are the initial states, and on the right side are
the mediate states. The process (a) leads to an effective attraction between
electron and hole, and the process (b) leads to an effective attraction
between electrons.}
\end{figure}

Usually the Hund coupling in the doped manganites is very strong, i.e., $%
J_{H}S\gg t$. Large $J_{H}S$ suggests that most electrons form spin $S+1/2$
states with the localized spins on the same sites, which makes it
appropriate to utilize the projective perturbation technique to investigate
the low-energy physics of the Hamiltonian (\ref{ele-ham}). The effect of
finite and large $J_{H}S$ can be regarded as the perturbation correction to
the case of infinite $J_{H}$, which is described by a quantum double
exchange model. \cite{Kubo72} Up to the second-order perturbation
correction, there are two types of the virtual processes which contribute to
the low energy physics (See in Fig. 1): (a). An electron hops from one site
to one of the nearest neighbor empty site to form a spin $S-1/2$ state and
then hops backward. The intermediate state has a higher energy $\Delta
E_{a}=J_{H}(S+1/2)$ than the initial state. (b). One electron hops from one
site to one of the singly occupied sites and then backward. The intermediate
state has a higher energy $\Delta E_{b}=J_{H}S+U$ than that of the initial
state. Hence, by using a projective perturbation approach, \cite
{Shen97,Shen98} the effective Hamiltonian is written as \cite{Note} 
\begin{eqnarray}
H_{eff} &=&-t\sum_{ij,\sigma }\bar{c}_{i,\sigma }^{\dagger }\bar{c}%
_{j,\sigma }+J_{AF}\sum_{\langle ij\rangle }\bar{{\bf S}}_{i}\cdot \bar{{\bf %
S}}_{j}  \nonumber \\
&+&\frac{2St^{2}}{J_{H}(2S+1)^{2}}\sum_{ij}\left( \frac{{\bf S}_{i}}{S}\cdot 
\frac{\tilde{{\bf S}}_{j}}{S+\frac{1}{2}}-1\right) P_{ih}P_{js}^{+} 
\nonumber \\
&+&\frac{t^{2}}{J_{H}S+U}\sum_{ij}\left( \frac{\tilde{{\bf S}}_{i}}{S+\frac{1%
}{2}}\cdot \frac{\tilde{{\bf S}}_{j}}{S+\frac{1}{2}}-1\right)
P_{is}^{+}P_{js}^{+}  \label{eff-ham}
\end{eqnarray}
where $\bar{{\bf S}}_{i}={\bf S}_{i}P_{ih}+2S\tilde{{\bf S}}%
_{i}P_{is}^{+}/(2S+1)$ and 
\[
\bar{c}_{i,\sigma }=\sum_{\sigma ^{\prime }}\frac{{\bf S}_{i}\cdot {\bf %
\sigma }_{\sigma \sigma ^{\prime }}+(S+1)\delta _{\sigma \sigma ^{\prime }}}{%
2S+1}(1-n_{i,-\sigma ^{\prime }})c_{i,\sigma ^{\prime }}. 
\]
$\tilde{{\bf S}}_{i}$ is a spin operator with spin $S+1/2$, and a
combination of spin of electron and localized spin on the same site. $P_{ih}$
and $P_{is}^{+}$ are the projection operators for empty site and single
occupancy of spin $S+1/2$. The first term in Eq.(\ref{eff-ham}) is the
quantum double exchange model. \cite{Kubo72,Shen97} It enhances
ferromagnetic correlation, and may be suppressed if the antiferromagnetic
exchange coupling of localized spin is very strong. The second, third and
fourth terms prefer antiferromagnetism to ferromagnetism. The third term
describes an attractive particle-hole interaction since the value of the
operator before $P_{ih}P_{js}^{+}$ is always non-positive. In another words,
an repulsive interaction between electrons in the restricted space arises
when the spin background deviates from a saturated ferromagnetic case.

To simplify the problem, we take the large spin approximation, and keep $%
J_{H}S=j_{h}$ and $J_{AF}S^{2}=j_{af}$. The spin operator is parameterized
in polar angle $\theta $ and $\phi $. In the approximation, the Hamiltonian
is further reduced to 
\begin{eqnarray}
H_{cl} &=&-t\sum_{ij}c_{ij}\alpha _{i}^{\dagger }\alpha
_{j}-2j_{af}\sum_{ij}\sin ^{2}\frac{\Theta _{ij}}{2}  \nonumber \\
&+&\sum_{ij}2\sin ^{2}\frac{\Theta _{ij}}{2}\left( \frac{t^{2}}{2j_{h}}-%
\frac{t^{2}}{j_{h}+U}\right) \alpha _{i}^{\dagger }\alpha _{i}\alpha
_{j}^{\dagger }\alpha _{j}  \nonumber \\
&-&\sum_{ij}\frac{t^{2}}{2j_{h}}\sin ^{2}\frac{\Theta _{ij}}{2}(\alpha
_{i}^{\dagger }\alpha _{i}+\alpha _{j}^{\dagger }\alpha _{j})  \label{cl-ham}
\end{eqnarray}
where 
\[
c_{ij}=\cos \frac{\theta _{i}}{2}\cos \frac{\theta _{j}}{2}+\sin \frac{%
\theta _{i}}{2}\sin \frac{\theta _{j}}{2}e^{-i(\phi _{i}-\phi _{j})} 
\]
; 
\[
\cos \Theta _{ij}=\cos \theta _{i}\cos \theta _{j}+\sin \theta _{i}\sin
\theta _{j}\cos (\phi _{i}-\phi _{j}); 
\]
\[
\alpha _{i}=\cos \frac{\theta _{i}}{2}(1-n_{i,\downarrow })c_{i,\uparrow
}+\sin \frac{\theta _{j}}{2}(1-n_{i,\uparrow })c_{i,\downarrow }. 
\]
Physically, $\alpha $ is an electronic operator which is fully polarized
along the localized spin on the same site. $|c_{ij}|=\cos (\Theta _{ij}/2)$
and approaches to zero when $\Theta _{ij}\rightarrow \pi $. If we neglect
the Berry phase in $c_{ij}$, the first term gets back to the classical DE
model. Now it is clear that the ferromagnetism is always predominant in the
ground state if other terms in the effective Hamiltonian (Eq. (\ref{cl-ham}%
)) are neglected. The sign of the interaction 
\begin{mathletters}
\begin{equation}
V_{ij}=2\sin ^{2}\frac{\Theta _{ij}}{2}\frac{t^{2}}{2j_{h}}\frac{U-j_{h}}{%
U+j_{h}}  \label{interaction}
\end{equation}
is determined by the ratio $j_{h}/U$. If $U$ is less than $j_{h}$, the
interaction is attractive, but if $U$ is greater than $j_{h}$, the
interaction is repulsive. The attractive or repulsive interaction will lead
to different physics. Hence $U=j_{h}$ is a quantum critical point. The
influence of the on-site Coulomb interaction will change qualitatively (not
just quantitatively) the physics of the doped manganites, which is usually
ignored. In the case of small $U$, the attractive interaction will drive
electrons to accumulate together to form an electron-rich regime. i.e., the
phase separation may occur when the spin background becomes
antiferromagnetism. \cite{Shen98} Monte Carlo simulation by Dagotto {\it et
al.} \cite{Dagotto97} shows that the phase separation occurs in the case of $%
U=0$. However, the phenomenon was not observed in the case of large $U$. The
phase diagram of $U=0$ is also seen in Ref.\cite{Riera97}. From our
analysis, the attractive interaction originates from the virtual process
(b). Due to the double occupancy in the intermediate state, an extra energy $%
U$ costs in the process. When $U$ is sufficiently large, the process (b)
will be suppressed and the process (a) becomes predominant. The net
interaction between electrons is repulsive. Therefore the phase separation
may occur only if $U<j_{h}$.

\section{Origin of Wigner lattice}

We are now\ in the position to discuss the instability to the Wigner
lattice. In the doped manganites, the on-site Coulomb interaction is much
stronger than the Hund's rule coupling, {\it i.e.}, $U\gg J_{H}S.$ \cite
{Satpathy96} In the case, the process (b) in Fig.1 needs a much higher
energy to be excited than the process (a) does. The process (a) dominates
over the process (b). The effective interaction is repulsive. Hence we shall
focus on the case of strong correlation (i.e., $U\gg J_{H}S$). To simplify
the problem, we take $U\rightarrow +\infty $ and neglect the term containing 
$U$ in Eq.(\ref{cl-ham}). A finite and large $U$ will produce minor
quantitative (not qualitative) changes of the physics we shall discuss. The
ratio of the repulsion to the hopping term $r=(t/j_{h})\sin ^{2}\frac{\Theta
_{ij}}{2}/\cos \frac{\Theta _{ij}}{2}$ depends on not only $t/j_{h}$, which
is usually very small, but also the angle of two spins. $r=0$ if $\Theta
_{ij}=0$, and $+\infty $ if $\Theta _{ij}=\pi $. In other words, the ratio
could become divergent in the antiferromagnetic spin background ($\Theta
_{ij}=\pi $) even though $t/j_{h}$ is very small. Relatively large ratio
will make a state with a uniform density of electrons unstable. To
understand the physical origin for the Wigner lattice at $x=0.5$, we first
see what happens in the antiferromagnetic background. When all $\Theta
_{ij}\rightarrow \pi $, the average energy per bond is $-2j_{h}$ if the two
sites are empty or occupied, and $-(2j_{af}+t^{2}/j_{h})$ if one site is
empty and another one is occupied. The later has a lower energy. At $x=1/2$, 
$\langle \alpha _{i}^{\dagger }\alpha _{i}\rangle =1/2$. The average energy
per bond is $-(2j_{af}+t^{2}/2j_{h})$ for a state with a uniform density of
electrons. If the electrons form a Wigner lattice, i.e., $\langle (\alpha
_{i}^{\dagger }\alpha _{i}-1/2)(\alpha _{j}^{\dagger }\alpha
_{j}-1/2)\rangle =-1/4$, the average energy per bond is $%
-(2j_{af}+t^{2}/j_{h})$, which is lower than that of the state with a
uniform density. Therefore in the antiferromagnetic background a uniform
density state is not stable against the Wigner lattice even for a small $%
t/j_{h}$. The same conclusion can be reached by means of the random phase
approximation. On the other hand, the formation of the Wigner lattice will
also enhance the antiferromagnetic exchange coupling from $-j_{af}$ to $%
-(j_{af}+t^{2}/2j_{h})$.

The phase diagram of the ground state is determined by the mean field
approach. Several of the features are determined in several limits: for
example, the ground state is ferromagnetic at $t/j_{h}=0$ and $j_{af}=0$.
Due to the instability to the Wigner lattice or charge density wave for
finite $t/j_{h}$ and $j_{af}$ we take $\langle \alpha _{i}^{\dagger }\alpha
_{i}-1/2\rangle =\Delta e^{i{\bf Q}\cdot {\bf r}_{i}}$ where ${\bf Q}=(\pi
,\pi ,\cdots )$ and $\langle \cdots \rangle $ is the ground state average.
We also take $\langle c_{ij}\rangle =\cos (\Theta /2)$ and $\langle \sin
^{2}(\Theta _{ij}/2)\rangle =\sin ^{2}(\Theta /2)$.\cite{Calderon98} The
free energy per bond is 
\end{mathletters}
\begin{eqnarray}
{\cal E}(\Delta ,\Theta ) &=&-\int \frac{dk}{(2\pi )^{d}}\sqrt{\epsilon
^{2}(k)\cos ^{2}\frac{\Theta }{2}+4\frac{t^{4}}{j_{h}^{2}}\sin ^{4}\frac{%
\Theta }{2}\Delta ^{2}}  \nonumber \\
&-&(j_{af}+\frac{1}{4}\frac{t^{2}}{j_{h}})\sin ^{2}\frac{\Theta }{2}+\frac{%
t^{2}}{j_{h}}\sin ^{2}\frac{\Theta }{2}\Delta ^{2}
\end{eqnarray}
where $\epsilon (k)=-t(\sum_{\alpha =1}^{d}\cos k_{\alpha })/d$ and d is the
number of dimension. The integration runs over the reduced Brillouin zone.
The phase diagram (Fig. 2) is obtained by minimizing the energy ${\cal E}%
(\Delta ,\Theta )$. $\Delta $ and $\Theta $ are the order parameters for
charge and magnetic orderings, respectively. $\Delta =0$ and $\Theta =0$
represents a full ferromagnetic (FM) phase, $\Delta =0$ and $\Theta \neq 0$
represents a canted ferromagnetic (CF) phase, $\Delta =1/2$ and $\Theta =\pi 
$ represents the Wigner lattice (WL), and $\Delta <1/2$ and $\Theta \neq 0$
represents a mixture of charge and spin density waves. A full ferromagnetic
phase diagram appears at smaller $t/j_{h}$ and $j_{af}$, which indicates
that the double exchange ferromagnetism is predominant. The Wigner lattice
appears at a larger $t/j_{h}$ and $j_{af}$. The antiferromagnetic coupling
originating from the virtual precess (a) and superexchange coupling of the
localized spins can suppress the double exchange ferromagnetism completely.
A canted ferromagnetic phase is between the two phases. At $j_{af}=0$, the
transition from ferromagnetism to the Wigner lattice occurs at $%
t^{2}/j_{h}=2\int dk\epsilon (k)/{(2\pi )^{d}}$ which equals to $0.63662t$
for $d=1$, $0.405282t$ for $d=2$, and $0.336126t$ for $d=3$. When the
effective potential energy $t/j_{h}$ becomes to dominate over the kinetic
energy, the ferromagnetic phase is unstable against the Wigner lattice. For
a finite $j_{af}$, a smaller $t/j_{h}$ is required to form a Wigner lattice.
However $t/j_{h}$ must be nonzero, even for a large $j_{af}$. In the double
exchange model, i.e. $j_{h}\rightarrow +\infty $, we do not expect that the
Wigner lattice could appear at low temperatures at $x=1/2$ unless a strong
long-range Coulomb interaction is introduced.

\begin{figure}[tbp]
\epsfxsize=8.5cm
\epsfbox{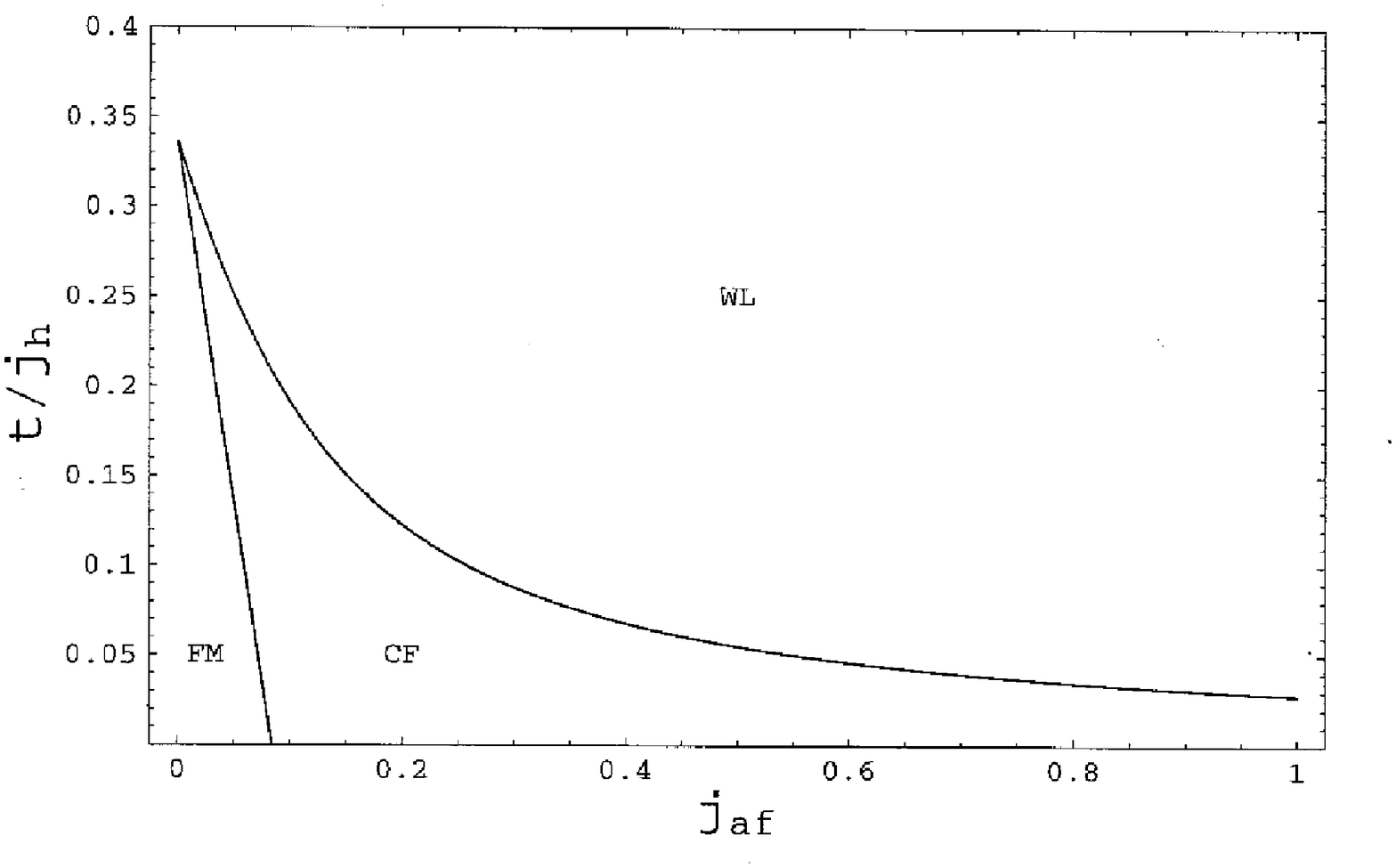}
\caption{ The phase diagram of the ferromagnetic Kondo lattice model on a
cubic lattice ($d=3$) at $x=1/2$.}
\end{figure}

\section{Ferrimagnetism and Wigner lattice}

We go back to Eq. (\ref{eff-ham}) to discuss the magnetic properties of the
ground state (or at zero temperature) in the case that the Wigner lattice is
formed at $x=1/2$ ($1-x$ is the density of electrons). The charge ordering
in the manganite is an alternating Mn$^{3+}$ and Mn$^{4+}$ arrangement
rather than a charge density modulation, which means $n_{i}=1$ or 0. A
d-dimensional hyper-cubic lattice can be decomposed onto two sublattice $%
{\cal A}$ and ${\cal B}$. In the charge ordering state, suppose that all
electrons occupy the sublattice ${\cal A}$, then 
\begin{equation}
P_{ih}P_{js}^{+}=\left\{ 
\begin{array}{ll}
1, & \makebox{if $i\in {\cal B}$ and $j\in {\cal A}$}; \\ 
0, & \makebox{otherwise.}
\end{array}
\right.
\end{equation}
The first term in Eq.(\ref{eff-ham}) must be suppressed completely as the
Wigner lattice is a static real space pattern, {\it i.e.}, the hopping
processes are forbidden. In the case, the Hamiltonian is reduced to 
\begin{equation}
H_{AF}=J_{AF}^{\prime }\sum_{i\in {\cal B},j\in {\cal A}}\left( \frac{{\bf S}%
_{i}}{S}\cdot \frac{\tilde{{\bf S}}_{j}}{S+\frac{1}{2}}-1\right)
\label{anti-ham}
\end{equation}
where $J_{AF}^{\prime }=J_{AF}S^{2}+2St^{2}/J_{H}(2S+1)^{2}$ and the
summation runs over the nearest neighbor pairs. This is an antiferromagnetic
Heisenberg model. The spin on the sublattice ${\cal A}$ is $S+1/2$ as the
electrons on the sites form spin $S+1/2$ state with the localized spins, and
the spin on the sublattice ${\cal B}$ is $S$. According to Lieb-Mattis
theorem, \cite{Lieb62} the ground state of Eq.(\ref{anti-ham}) is unique
apart from spin SU(2) ($2S_{tot}+1$)-fold degeneracy. The total spin of the
ground state $S_{tot}$ is equal to the difference of the maximal total spins
of two sublattices. In the case, 
\begin{equation}
S_{tot}=\frac{N_{e}}{2}
\end{equation}
which is also the maximal total spin of electrons ($N_{e}$ is the number of
electrons). It seems to be that all electrons are saturated fully while the
localized spins form a spin singlet state. Furthermore, it is shown
rigorously that the ground state possesses antiferromagnetic long-range
order as well as ferromagnetic one for any dimension. \cite{Shen94}

\section{Discussion and summary}

We wish to point out that, in the case that the Wigner lattice is formed,
the magnetic structure established here is unlikely in full agreement with
all experimental observations. \cite{Wollan55,Tokura96} The model discuss
here is a simplified theoretical model which has neglected some effects,
such as the orbital degeneracy of $e_{g}$ electrons, strong John-Teller
effect and lattice distortion. Methodologically, we apply the projective
perturbation approach to deal with the model. The strong electron-electron
correlations has been successfully taken into account by the projection
process. The perturbation process tells us that the effective Hamiltonian
should be valid at small $t/j_{h}$, which requires a strong Hund coupling
comparing with the hopping integral $t$. In practice, the parameters of the
model for doped manganites are roughly estimated as $U\approx 5.5eV$, $%
J_{H}\approx 0.76eV$, $t\approx 0.41eV$, $J_{AF}\approx 2.1meV.$\cite
{Satpathy96} Thus,\ $U/J_{H}S\approx 4.82$ and $\ t/J_{H}S\approx 0.359$.
For these parameters, the Wigner lattice at low temperatures is stable in
the phase diagram in Fig.2. Therefore, the superexchange process in Fig.1(a)
should play an important role in driving electrons to form the Wigner
lattice no matter whether the direct nearest neighbor Coulomb interaction is
strong. It is worth mentioning that  the direct Coulomb interaction will
always favor to form the Wigner lattice. \cite{Lee97} If the direct Coulomb
interaction is also included in the electronic model, which is not much
screened, the stability of Wigner lattice will be greatly enhanced. Note
that the Coulomb interaction is independent of the magnetic structure, and
should not be very sensitive to an external magnetic field. The effect of
field-induced melting of the Wigner lattice suggests that the physical
origin of the state may be closely related to the magnetic structure, which
is an essential ingredient of the present theory. In the actual compounds,
both the mechanisms should have important impact on the electronic
behaviors. It is unlikely that only one of them is predominant. As for the
mean field approximation, when it is sure that the instability of Wigner
lattice occurs at low temperatures, it is an efficient and powerful tool to
determine the phase diagram, although some other physical quantities, such
as critical exponents, cannot be obtained accurately. Due to the strongly
correlations of electrons, it is still lack of numerical results to verify
the present theoretical prediction as this is the first time to discuss
instability of the Wigner lattice in a model without nearest neighbor or
long-range interactions. When the system is deviated from $x=0.5$, the
superexchange interaction is still very important to determine the behaviors
of electrons. Recently, it was observed that the charge stripes in (La,Ca)MnO%
$_{3}$ pair. \cite{Mori98} However, the two pairing stripes of Mn$^{3+}$
ions are separated by a stripe of Mn$^{4+}$ ions. This fact suggests the
nearest neighbor interaction should be very strong. Of course, for a
comprehensive understanding of the phase diagram, including anisotropic
properties of charge and magnetic orderings, we need to take other effects
into account.

The role of Hund's rule coupling in the doped manganites has been emphasized
since the double exchange mechanism was proposed. However, the rich phase
diagrams in the doped manganites go beyond the picture. Our theory shows
that the on-site Coulomb interaction also has an important impact on the
physical properties of the system. In the model we investigate, the sign of
the effective interaction in Eq.(\ref{interaction}) depends on the ratio of j%
$_{h}$/U. Repulsive or attractive interaction will lead to quite different
physics. In one of our recent papers \cite{Shen98}, we proposed a mechanism
of phase separation based on the attractive interaction caused by the
virtual process (a) in Fig. 1, and neglect the on-site interaction U. The
phase separation can occur in the high and low doping regions. As the
mechanism of the phase separation is completely opposite to the mechanism of
the Wigner lattice we discuss in this paper, we have to address the issue
which one \ occurs for the doped manganites. From the estimation of the
model parameters for the actual compound, $U/j_{h}\approx 4.82$. Thus, the
effective interaction should be repulsive, not attractive. From this sense,
the phase separation we predicted in Ref. \cite{Shen98} could not occur in
doped manganites. In fact, both the phase separation and the Wigner lattice
were observed in the family of samples with different dopings. For example
the phase separation was observed in La$_{1-x}$Cu$_{x}$MnO$_{3}$ with x=0.05
and 0.08.\cite{Allodi97} It is worth pointing out that the electronic model
is a simplified model for doped manganites since the degeneracy of e$_{g}$
electrons and the Jahn-Teller effect have been neglected. The importance of
the orbital degeneracy of e$_{g}$ electron has been extensively discussed,
especially for the ferromagnetism near $x=0$. If we take into account the
orbital degeneracy, there may exist an superexchange virtual process in the
ferromagnetic or A-type antiferromagnetic background, in which the
superexchange coupling between different orbits instead of the spin indices
in Fig. 1 could produce an attractive interaction as we predicted in Ref. 
\cite{Shen98}. The mechanism for phase separation may still be responsible
for the experimental observation. The investigation along this direction is
in progress.

Before ending this paper, we would like to address the stability of the
Wigner lattice with respect to the transfer $t$. Some experimental analysis
suggested that a {\it relatively small} $t$ would favor to form the Wigner
lattice,\cite{Tokura96} which seems to be unlikely in contradiction with the
phase diagram in Fig. 2. In the present theory, the Wigner lattice occurs in
a moderate value of $t$. On one hand, a large $t$ ($\gg j_{h}$), of course,
will lead to the instability of the Wigner lattice and destroy the double
exchange ferromagnetism. In the case, a paramagnetic phase should be favored
at low temperatures. The perturbation technique used in this paper is also
not valid. So the region of Wigner lattice in Fig.2 cannot be naively
extended to the large $t$ case. On the other hand, when $t$ becomes very
small comparing with $j_{h}$, the Wigner lattice should also be unstable
since a small t means to enhance the ratio $j_{h}/t$ and a larger ratio is
favorable to double exchange ferromagnetism. If the antiferromagnetism from $%
t_{2g}$ electrons could compete over the double exchange ferromagnetism at $%
x=0.5$, it would suppress ferromagnetism at all the range of $x$.\cite
{Shen97} The effective transfer $t$cos$(\Theta /2)$ is determined by either
t or $\Theta $ the angle of the two spins. The Wigner lattice is also
accompanied by the strong antiferromagnetic correlation. The field-induced
melting effect indicates that the Wigner lattice is unstable in the
ferromagnetic background, which also indicates the important role of the
antiferromagnetic correlation to stabilize the Wigner lattice. A smaller $%
j_{af}$ will reduce the angle $\Theta $ and should also lead to the
instability of the Wigner lattice. Thus, a small $t$ does not always favor
to form the Wigner lattice.

In short, we derived an effective Hamiltonian for an extended Kondo lattice
model, based on which a physical mechanism for charge ordering in half-doped
manganites is naturally put forward.

\acknowledgments       
This work was supported by a CRCG research grant at the University of Hong
Kong.


\begin{references}
\bibitem{Note}  We neglect a three-site term which describes the indirect
hopping between the next nearest neighbour sites and is quite smaller than
the direct hopping term.
\end{references}
\end{document}